\documentclass[
 prx,
 reprint,
 superscriptaddress,
 showpacs,preprintnumbers,
nofootinbib,
 amsmath,amssymb,
 aps,
]{revtex4-1}

\usepackage{graphicx}
\usepackage[]{textcomp}
\usepackage[]{xspace}
\usepackage{dcolumn}
\usepackage{color}
\usepackage{color,soul}
\usepackage{graphicx}
\usepackage{dcolumn}
\usepackage{bm}
\usepackage{physics}
\usepackage[T1]{fontenc} 
\usepackage{soul}

\begin{document}

\title
 {Phonon-induced multi-color correlations in hBN single-photon emitters}

\author{Matthew A. Feldman}
\email{Matthew.Feldman@vanderbilt.edu}
\affiliation{Department of Physics and Astronomy, Vanderbilt University, Nashville, TN 37235, USA}
\affiliation{Quantum Information Science Group, Oak Ridge National Laboratory, Oak Ridge, TN 37831, USA}
\author{Alex Puretzky}
\affiliation{Center for Nanophase Materials Sciences, Oak Ridge National Laboratory, Oak Ridge, TN 37831, USA}
\author{Lucas Lindsay}
\affiliation{Materials Science and Technology Division, Oak Ridge National Laboratory, Oak Ridge, TN 37831, USA}
\author{Ethan Tucker}
\affiliation{Quantum Information Science Group, Oak Ridge National Laboratory, Oak Ridge, TN 37831, USA}
\author{Dayrl P Briggs}
\affiliation{Center for Nanophase Materials Sciences, Oak Ridge National Laboratory, Oak Ridge, TN 37831, USA}
\author{Philip G. Evans}
\affiliation{Quantum Information Science Group, Oak Ridge National Laboratory, Oak Ridge, TN 37831, USA}
\author{Richard F. Haglund}
\affiliation{Department of Physics and Astronomy, Vanderbilt University, Nashville, TN 37235, USA}
\author{Benjamin J. Lawrie}
\email{lawriebj@ornl.gov}
\affiliation{Quantum Information Science Group, Oak Ridge National Laboratory, Oak Ridge, TN 37831, USA}
\date{\today}
\begin{abstract}
Color centers in hexagonal boron nitride have shown enormous promise as single-photon sources, but a clear understanding of electron-phonon interaction dynamics is critical to their development for quantum communications or quantum simulations. We demonstrate photon antibunching in the filtered auto- and cross-correlations $g^{(2)}_{lm}(\tau)$ between zero-, one- and two-phonon replicas of defect luminescence. Moreover, we combine autocorrelation measurements with a violation of the Cauchy-Schwarz inequality in the filtered cross-correlation measurements to distinguish a low quantum-efficiency defect from phonon replicas of a bright defect. With no background correction, we observe single photon purity of $g^{(2)}(0)=0.20$ in a phonon replica and cross-spectral correlations of $g^{(2)}_{lm}(0)=0.18$ between a phonon replica and the zero phonon line. These results illustrate a coherent interface between visible photons and mid-infrared phonons and provide a clear path toward control of photon-phonon entanglement in 2D materials.
\end{abstract}


\maketitle

The recent discovery of a wide class of defect-based single photon emitters (SPEs) in hexagonal boron nitride (hBN) has spurred significant interest in the development of two-dimensional (2D) materials and van der Waals heterostructures~\cite{Tran2015,Tran2016,Exarhos2017,Grosso2017,Sontheimer2017,Kianinia2017}. Defects in hBN have narrow linewidths~\cite{Sontheimer2017}, bright emission~\cite{Tran2015}, small Huang-Rhys factors~\cite{Tran2015,Exarhos2017}, and are stable at temperatures as high as 800K~\cite{Kianinia2017}. Stable SPEs in hBN thus far have only been categorized phenomenologically into two groups on the basis of the phononic contributions to their spectra. Group I color centers have an asymmetric zero-phonon-line (ZPL) sideband~\cite{Anonymous:2016da,Tran2016} and a doublet optical phonon sideband redshifted $\sim160 (5)$ meV from the ZPL~\cite{Tran2015,Anonymous:2016da,Martinez2016,Shotan2016,Chejanovsky2016,Tran2016,Kianinia2017}. Group II defects have a symmetric ZPL and less pronounced optical phonon sidebands~\cite{Tran2016}.

The state structure of hBN defects, the large variance in ZPL energies, and the electron-phonon dynamics and energetics remain poorly understood. Improved understanding of these properties will drive the development of 2D hybrid quantum systems that leverage quantum coherent photonic and phononic interactions to generate indistinguishable single photons and to enable quantum frequency conversion~\cite{Vuong2016}. Coherent phonon-emitter coupling has been explored in depth for diamond color centers, resulting in the observation of phonon-mediated photon bunching~\cite{feldman2018colossal}, quantum teleportation from photonic to phononic states~\cite{hou2016quantum}, phononic quantum memories suitable for storing single photons~\cite{england2013photons,england2015storage}, and room-temperature phononic quantum processing~\cite{lee2012macroscopic}.

In this Letter we explore the electron-phonon dynamics of group I defects in few-layer hBN with one- and two-color Hanbury Brown-Twiss (HBT) interferometry and micro-photoluminescence ($\mu$PL) spectroscopy. We provide the first evidence of single-phonon excitation and photon-phonon entanglement in 2D materials by measuring antibunching in the one- and two-phonon replicas of a hBN color center and we use violations of the Cauchy-Schwarz inequality in two-color cross-correlation measures to distinguish low quantum-efficiency defects from phonon replicas of bright defects. These measurements demonstrate that hBN is an ideal platform for generating single mid-infrared phonons by optical excitation because of the weak electron-phonon coupling in hBN compared with diamond. This is a critical step toward the realization of deterministic single-phonon sources and acoustic quantum transducers that have been proposed in recent years~\cite{schuetz2015universal,sollner2016deterministic}.

\begin{figure}[]
\centering
    \includegraphics[width=\columnwidth]{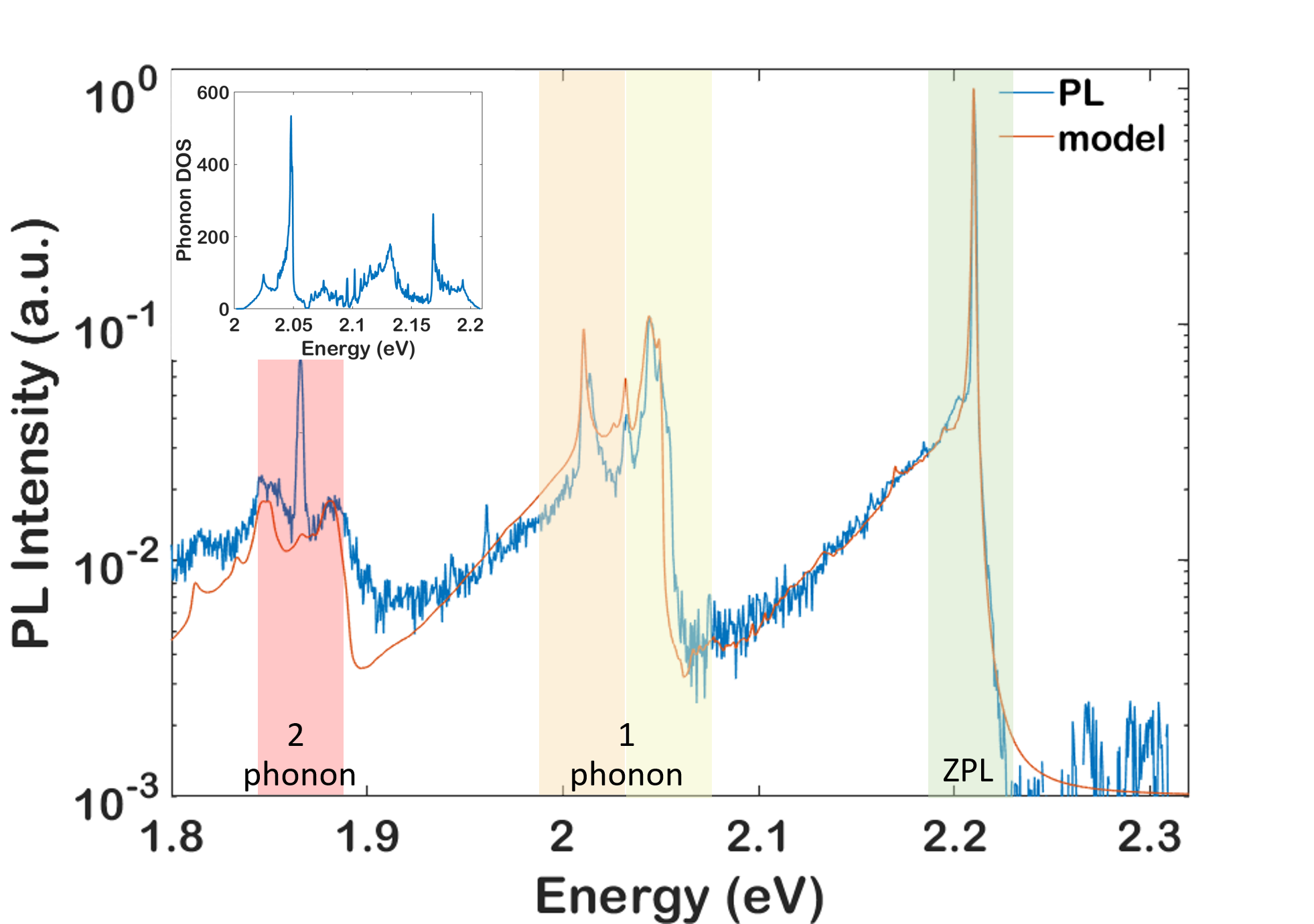}
    \caption{Background corrected defect $\mu$PL spectrum (blue) collected at 3.6K rescaled by $1/\omega^3$ for direct comparison with the calculated $\mu$PL cross-section . The calculated one-phonon density of states (DOS) is shown in the inset and the $\mu$PL spectrum calculated from the rescaled phonon DOS is shown in orange. The calculated $\mu$PL provides a reasonable reproduction of the ZPL and the one-phonon replicas, but an additional transition appears in the experimental $\mu$PL spectrum that is not present in the calculated two-phonon replicas.}
    \label{fig:fig1}
\end{figure}

\begin{figure*}[]
\centering
    \includegraphics[width=0.75\textwidth]{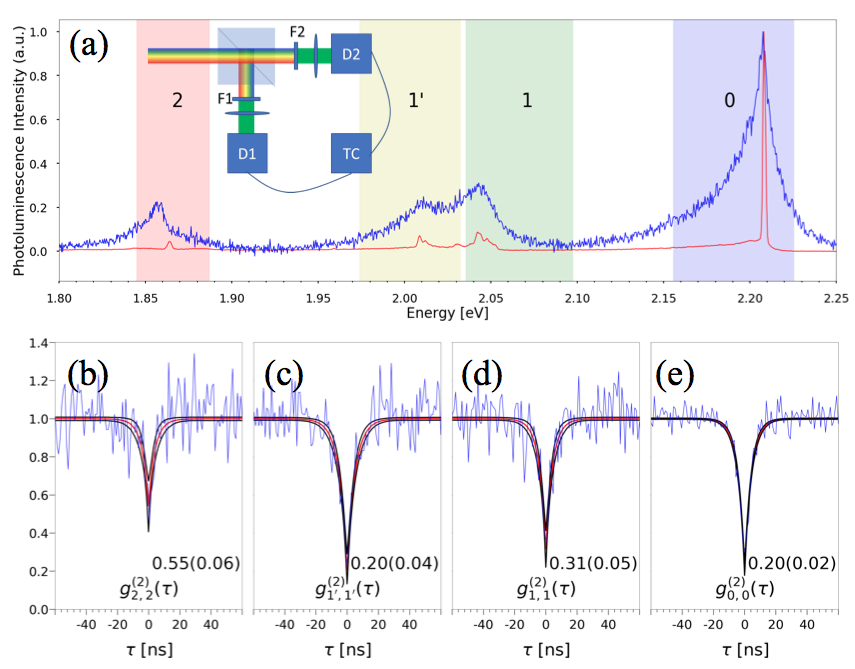}
    \caption{(a) Background corrected room temperature (blue) and 3.6K (red) PL spectrum of a hBN defect (the inset illustrates a HBT interferometer with two single-photon detectors (D1 and D2), tunable bandpass filters (F1 and F2), and high speed time correlation electronics (TC)). The filters are actively tunable from 1.75-2.95 eV, enabling spectrally-resolved single photon detection across the bandwidth of all observed SPEs. (b-e) Frequency-filtered two-photon autocorrelations $g^{(2)}_{ll}(\tau)$  (blue), median fits (red), and 95$\%$ credibility intervals (black) for the two-phonon replicas (2), LO($\Gamma$) and LO(T) replicas (1$^\prime$), TO(M)/LO(K) replicas (1), and ZPL (0), respectively. The colored bands (0,1$^\prime$,1,2) in (a) are the spectral ranges of the bandpass filters used for each of the autocorrelation measurements in (b-e). The mean and standard deviation for $g^{(2)}_{ll}(0)$ are inset in panel (a)-(e).}
    \label{fig:fig2}
\end{figure*}

We used a custom confocal microscope to excite color centers in hBN with a 405 nm CW source. Nineteen group I defects with ZPLs varying from 1.7-2.7 eV were surveyed~\cite{suppmatt} with $\mu$PL and HBT interferometry, and a single defect exhibiting minimal coupling to other defects was chosen for more detailed spectroscopy. Background corrected $\mu$PL spectra were collected at temperatures of 3.6K and 300K, photon antibunching was measured for each transition identified in the $\mu$PL spectra, and photon cross-correlations were measured for each pair of transitions. Further details on the experimental apparatus are available in the supplemental material~\cite{suppmatt}. As seen in Fig~\ref{fig:fig1} and Fig~\ref{fig:fig2}(a), the group I defect examined here has a ZPL at 2.21 eV with a linewidth that broadens from 1.3 to 20 meV with increasing temperature, consistent with a temperature-dependent Debye-Waller factor associated with acoustic phonons~\cite{Anonymous:2016da}. hBN defects with ZPLs near 2.21 eV remain poorly classified\cite{weston2018native}, but the results discussed here are relevant to broad classes of defects in hBN. Phonon sidebands were observed redshifted by 166, 177, 200, 326, 343, 359 and 395 meV from the ZPL in agreement with Raman spectroscopy, inelastic x-ray scattering experiments and ab initio calculations of the phonon dispersion for hBN~\cite{Serrano2007,Kern1999,Geick1966,Nemanich1981,jiang2018anisotropic}.

In order to explain the vibronic structure of a specific defect, both localized defect vibrations and delocalized lattice vibrations should be considered. Although lattice phonons are well defined, the defect vibrational modes that are determined by defect structure and composition are not. Even in the case of known defects, such calculations are challenging even for hBN monolayers~\cite{Tawfik2017}.  Here we will consider a limiting case assuming that the main contribution to the vibronic spectrum is given by lattice phonons. As we show below, this assumption provides a reasonably good agreement  between the measured and calculated spectra.The inset of Fig.~\ref{fig:fig1} illustrates the calculated one-phonon hBN density of states (DOS). The TO(M)/LO(K) (166 meV) and LO(T) (177 meV) phonon modes in the phonon DOS are also well-defined in the measured PL spectrum. The measured single-phonon replica 200 meV to the red of the ZPL can be described as a result of a Fr\"{o}hlich interaction that scales inversely with the phonon wavevector so that even in the case of small single-phonon DOS, a diverging electron-phonon matrix element results in a significant LO($\Gamma$) phonon replica at 200 meV~\cite{Vuong2016}. Two-phonon replicas were observed redshifted from the ZPL by 326 meV (2TO(M)/2LO(K)), 343 meV (LO(T) + TO(M)/LO(K)), 359 meV (LO($\Gamma$) +TO(M)/LO(K)) and 395 meV (2LO($\Gamma$)).


In order to better validate this description of the $\mu$PL spectrum, we modeled each phonon replica in terms of a change in vibronic state at a rate modeled by Fermi's golden rule. We describe excited  $\widetilde{i}=| E,m\rangle$  and ground $\widetilde{f}=| G,n\rangle$ vibronic states, where m and n are the vibrational states of the lattice, and the electronic dipole operator $\mu$ is assumed to be independent of the nuclear coordinates. The emission spectrum cross-section is then proportional to the transition rate from  $| E,m\rangle$  to $| G,n\rangle$ 
\begin{equation}
\sigma(\omega) = \frac{16\pi^2 c}{\hbar\omega}  |\langle\widetilde{f}| \mu | \widetilde{i} \rangle|^2 \delta(\omega_{\widetilde{f}\widetilde{i}}-\omega)\\,
\end{equation}
which can be approximated to first order as
\begin{equation}
\frac{8 \pi c|\mu|^2}{ \hbar \omega }e^{-S} \int_{-\infty}^{\infty}dte^{it(\omega_{GE}-\omega)}e^{S\zeta(t)},
\end{equation}
where $\zeta(t)$ is given by
\begin{equation}
\int d\Omega(\rho(\Omega)/\Omega^2)[(n(\Omega)+1)e^{i\Omega t}+n(\Omega)e^{-i\Omega t}],
\end{equation}
where $\omega$ is the emission frequency, $\omega_{EG}$ is the frequency of the electronic transition, $\Omega$ is the vibrational frequency, $H_G$ and $H_E$ are the nuclear Hamiltonians for point defects in the ground and excited states, and  $\mu$ is the matrix element of the electric dipole operator of the point defect~\cite{Davies1974,Maradudin1967,stoneham2001}. S is the Huang-Rhys factor, $n(\Omega)$ the thermal average number of phonons and $\rho(\Omega)$ the total density of phonon states. For our calculation of the emission spectrum, the empirically determined Huang-Rhys factor (S), a phenomenological term accounting for the observed acoustic phonons~\cite{Exarhos2017,Vuong2016}, and the calculated one-phonon density of states shown in the inset of Fig~\ref{fig:fig1} were used to approximate $\rho(\omega)$. Based on experimental observation, we assumed that the TO(M)/LO(K) (166 meV), LO(T) (177 meV), and LO($\Gamma$) (200meV) modes were the dominant one-phonon modes. Hence, the calculated one-phonon density of states was re-weighted using three Lorentzians centered at each of these one-phonon modes. The calculated emission spectrum, plotted in Fig~\ref{fig:fig1}, reproduces all of the essential features of the experimental spectrum except for the measured feature 343 meV redshifted from the ZPL. Further details on the emission spectrum cross-section may be found in the supplemental material \cite{suppmatt}.

For a single defect, the selection rules determined by $\mu$ allow transitions of only one electron from $| E,m\rangle$ to $| G,n\rangle$ within the coherence lifetime of the electron. Hence this indicates that all transitions $| E,m\rangle$ to $| G,n\rangle$ should be strongly anti-correlated, while a mid-infrared phonon should be strongly correlated with each phonon replica.  However, background luminescence, uncorrelated color centers, and incoherent coupling dynamics can all suppress the expected correlations. Hence, we next investigate the correlation dynamics experimentally using colored-HBT interferometry to measure $g_{lm}^{(2)}(\tau)$ for phonon mediated transitions $l$ and $m$.

For the colored HBT experiments, tunable bandpass filters with bandwidth of 20 nm were used to select the ZPL, one- and two-phonon sidebands. While low-temperature PL was used to map the phonon replicas to the phonon DOS in Fig.~\ref{fig:fig1}, all auto- and cross-correlation measurements employed room-temperature PL to maximize photon collection efficiency and minimize the excitation spot size for spatially-selective defect excitation. Figure~\ref{fig:fig2}a illustrates the room-temperature photoluminescence with colored bars 0, 1, 1$^\prime$, and 2 representing the spectral filters that were used to select each transition. These filter bands correspond to integrating across the ZPL and acoustic phonon modes, the TO(M)/LO(K) mode, the LO($\Gamma$) and LO(T) modes, and the above described two-phonon modes, respectively. The filtered two-photon autocorrelations $g^{(2)}_{ll}(\tau)$ in Fig.~\ref{fig:fig2}(b)-(e) for the $l=$ 0, 1, $1^\prime$, and 2 spectral bands confirm that antibunching is present in every phonon replica in addition to the ZPL, even without background correction. 

Mean fits and 95$\%$ credibility intervals from a self-consistent Bayesian regression for a two-level model of $g_{ll}^{(2)}(\tau)$ are plotted in red and black respectively. The autocorrelations involving the ZPL, TO(M)/LO(K), and LO(T) modes exhibit $g_{ll}^{(2)}(0)<0.5$, clearly demonstrating single-photon emission. The two-phonon replicas reveal $g_{ll}^{(2)}(0)=0.55\pm 0.06$, demonstrating that the band centered at 1.87 eV likely includes two transitions. The time constants for all of the autocorrelations in Fig.~\ref{fig:fig2} are $\tau \approx 4$ ns, indicating that details of the phonon-defect coupling are not critical to the dynamics of the phonon replicas themselves.

\begin{figure}[]
\centering
    \includegraphics[width=\columnwidth]{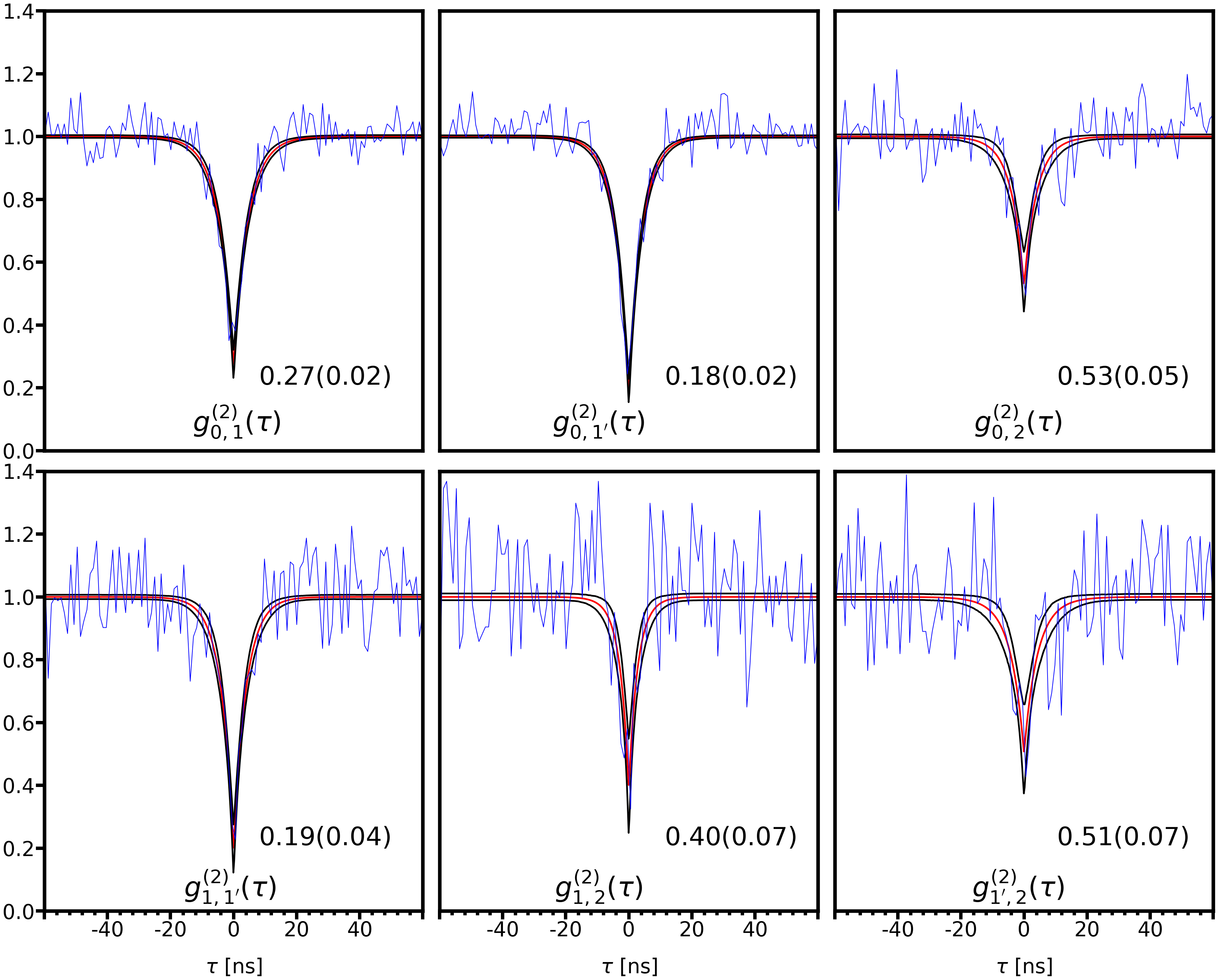}
    \caption{Frequency-filtered two-photon cross-correlations $g^{(2)}_{lm}(\tau)$ between each pair of spectral bands visible at room temperature, labeled in terms of the four spectral bands illustrated in Fig. 2(a).The mean and standard deviation for $g^{(2)}_{lm}(0)$ are inset in each panel}
    \label{fig:fig3}
\end{figure}

The frequency-filtered two-photon correlation function $g^{(2)}_{lm}(\tau)$ is a normalized measure of photon fluctuations that quantifies the correlation between a photon of color $l$ detected at time $\tau$ after a photon of color $m$ is detected. While Fig.~\ref{fig:fig2}(b)-(e) reported autocorrelations using the same bandpass filter on both detectors, Fig.~\ref{fig:fig3} uses different bandpass filters on each detector to measure the cross-correlations between all combinations of the four transitions. The same self-consistent Bayesian regression used to fit the autocorrelation functions in Fig.~\ref{fig:fig2}(b)-(e) is used in Fig.~\ref{fig:fig3}. 


The Cauchy-Schwarz inequality: $[g^{(2)}_{l,m}(\tau)]^2 \leq g^{(2)}_{l,l}(\tau)g^{(2)}_{m,m}(\tau)$ describes the classical limit for two mode fields.  While photon anticorrelations lead to the antibunching reported in Fig.~\ref{fig:fig2}, Cauchy-Schwarz inequality violations emerge from positive quantum correlations between two fields~\cite{marino2008violation,lee2012macroscopic,clauser1974experimental}. The Franck-Condon model of a single defect coupled to several phonon modes would lead one to expect anticorrelations between each measured frequency band, with no Cauchy-Schwarz inequality violation. Cascaded photoemission would yield a violation, as would the presence of uncorrelated single photon emitters within one of the spectral bands illustrated in Fig.~\ref{fig:fig2}(a).  All of the measured cross-correlations reported in Fig.~\ref{fig:fig3} satisfy the Cauchy-Schwarz inequality except for $g^{(2)}_{0,2}$ and $g^{(2)}_{1',2}$. Combining this observation with the reduced antibunching seen in the two-phonon replicas in Fig.~\ref{fig:fig2} and the phonon density of states calculations in Fig.~\ref{fig:fig1} provides significant support for the claim that the narrow linewidth feature seen at 1.86 eV in the low-temperature PL spectrum in Fig.~\ref{fig:fig1}(a) is an uncorrelated SPE. Note that no Cauchy-Schwarz violation is seen for $g^{(2)}_{1,2}$ because of the reduced antibunching of $g^{(2)}_{1,1}$ compared with $g^{(2)}_{0,0}$ and $g^{(2)}_{1',1'}$.

Distinguishing between low-brightness SPEs and phonon replicas is important for the further development of 2D materials for quantum technologies, but the critical issue here is the otherwise strong anticorrelations between antibunched spectral bands.  Because of the measured anticorrelations in Figs.~\ref{fig:fig2} and ~\ref{fig:fig3}, and because the measured PL spectrum can be completely described in terms of a ZPL and one and two phonon replicas, the state of the photoluminescence can be represented as $\alpha_1\ket{\psi_{\omega_{ZPL}}}+\sum_{i}\alpha_i\ket{\psi_{\omega_{PR_i}}}\ket{\psi_{\omega_{PH_i}}}+\sum_{j,k}\alpha_{j,k}\ket{\psi_{\omega_{PR_{j,k}}}}\ket{\psi_{\omega_{PH_j}}}\ket{\psi_{\omega_{PH_k}}},$ where each term describes a Fock state of zero or one photons or phonons. The first term describes photons emitted into the ZPL ($\omega_{ZPL}$), the second describes single phonon replicas ($\omega_{PR_i}$) and single mid-infrared phonons ($\omega_{PH_i}$), and the final term describes the two-phonon replica ($\omega_{PR_{j,k}}$) and the associated phonon pair($\omega_{PH_j}, \omega_{PH_k}$). Increased control over vibronic pathways will be critical to the generation of high fidelity photon-phonon entanglement, which may be enabled by appropriate phononic cavity design. The anticorrelations measured here thus provide a clear path toward heralded single phonon sources and phonon-photon entanglement in 2D materials.

The characterization of photonic correlations between all electronic and vibronic transitions associated with a given defect is critical to the understanding of quantum phononic dynamics. Further control of the phonon density of states is needed to generate controllable quantum vibronic states, but the measurements reported here have importance both for quantum information science and phononic technologies more generally. In particular, these results point toward one approach for developing quantum photonics and quantum phononics in the mid-infrared atmospheric transparency windows. Recent demonstrations of satellite-to-ground quantum key distribution, for example, have relied on weak coherent pulses and entangled single photon sources at near-infrared wavelengths~\cite{liao2017satellite,yin2017satellite}.  These demonstrations have suffered from 20-30 dB attenuation because of atmospheric absorption and scattering.  Developing bright single photon sources and entangled photon sources at mid-infrared wavelengths would enable significant improvements in satellite-to-ground quantum communications. 

Moreover, strong cross-correlations between the zero-phonon mode and the phonon replicas and strong positive correlations between each phonon replica and the associated mid-infrared phonon may enable side-channel attacks against quantum communication protocols. Similarly, detection of mid-infrared phonons could enable quantum non-demolition measurements on the state of visible phonon replicas. Because electron-phonon coupling in hBN is strain-dependent, the strength of these effects can be controlled with nanopatterned surfaces.

Finally, hexagonal boron nitride has drawn significant interest because of its hyperbolic phononic dispersion~\cite{li2015hyperbolic}.  Hyperbolic dispersion enables super-resolution imaging and significant Purcell effects for shallow defects.  Combining the single and entangled phonon sources described here with control over the hyperbolic dispersion of nanopatterned multilayer hBN could therefore enable significant advances in the concept of cavity quantum phonodynamics~\cite{soykal2011sound}. While the results presented here have illustrated the first evidence of quantum phononics in hBN, it remains crucial to more carefully explore the combined phononic, photonic, and electronic dynamics of color centers in 2D materials in order to advance science and technology in all of these research agendas.

\begin{acknowledgments}

The authors acknowledge feedback from Raphael Pooser. This research was sponsored by the Laboratory-Directed Research and Development Program of Oak Ridge National Laboratory, managed by UT-Battelle, LLC for the U.S. Department of Energy. M.F. gratefully acknowledges support by the Department of Defense (DoD) through the National Defense Science \& Engineering Graduate Fellowship (NDSEG) Program. L.L. acknowledges support from the U. S. Department of Energy, Office of Science, Basic Energy Sciences, Materials Sciences and Engineering Division. Rapid thermal processing and spectroscopy experiments were carried out at the Center for Nanophase Materials Sciences (CNMS), which is sponsored at ORNL by the Scientific User Facilities Division, Office of Basic Energy Sciences, U.S. Department of Energy.
\end{acknowledgments}

\end{document}